\documentclass[manuscript]{aastex63}
\accepted{\today}
\shorttitle{ }
\shortauthors{Reetika Joshi et al.}
\graphicspath{{./}{figures/}}

\begin{document}

\title{Cause and Kinematics of a Jet--Like CME}

\correspondingauthor{Reetika Joshi}
\email{reetikajoshi.ntl@gmail.com}
\author[0000-0003-0020-5754]{Reetika Joshi}
\affiliation{Department of Physics, DSB Campus, Kumaun University, Nainital -- 263 001, India}

\author[0000-0002-8887-3919]{Yuming Wang}
\affiliation{CAS Key Laboratory of Geospace Environment, School of Earth and Space Sciences, University of Science and Technology of China, Hefei, China}
\affiliation{CAS Center for Excellence in Comparative Planetology, University of Science and Technology of China, Hefei 230026, China}

\author[0000-0002-3518-5856]{Ramesh Chandra}
\affiliation{Department of Physics, DSB Campus, Kumaun University, Nainital -- 263 001, India}

\author{Quanhao Zhang}
\affiliation{CAS Key Laboratory of Geospace Environment, School of Earth and Space Sciences, University of Science and Technology of China, Hefei, China}

\author{Lijuan Liu}
\affiliation{School of Atmospheric Sciences, Sun Yat-sen University, Zhuhai, Guangdong, 519000, China}

\author{Xiaolei Li}
\affiliation{CAS Key Laboratory of Geospace Environment, School of Earth and Space Sciences, University of Science and Technology of China, Hefei, China}

\begin{abstract}

In this article, we present the multi--viewpoint and multi--wavelength analysis of an atypical solar jet based on the data from {\it Solar Dynamics Observatory}, {\it SOlar and Heliospheric Observatory}, and {\it Solar TErrestrial RElations Observatory}. It is usually believed that the coronal mass ejections (CMEs) are developed from the large scale solar eruptions  in the lower atmosphere. However, the kinematical and spatial evolution of the jet on 2013 April 28 guide us that the jet was clearly associated with a narrow CME having a width of $\approx$ 25$^\circ$ with a speed of $\approx$ 450 km s$^{-1}$. To better understand the link between the jet and the CME, we did the coronal potential field extrapolation from  the  line  of  sight magnetogram  of  the AR. The extrapolations present that the jet eruption follows exactly the same path of the open magnetic field lines from the source region which provides the route for the jet material to escape from  the solar surface towards the outer corona.

\end{abstract}


\section{Introduction} \label{sec:intro}

Solar jet is a common phenomenon of small scale plasma ejections from the solar lower atmosphere towards the solar corona. Solar flare or the base brightening at the jet footpoint is believed to promote the force for pushing the plasma material upward. After the first observational evidence of solar jets by \citet{Shibata1992} using Yohkoh satellite observations in X-rays, jets now become a popular research topic in solar physics.  They are studied by many authors \citep{Canfield1996, Alexander1999, Nistico2009, Filippov2009, Hong2011, Jiajia2014, Schmieder2013, Sterling2015,Jiajia2016,Chandra2017, Kayshap2018, Jiajia2019, Ruan2019}. From the previous reported results, it is now well accepted  that 68 $\%$ of solar jets are active region (AR) jets \citep{Shimojo1996,Sterling2017} and the length, velocity and width have an average value of $\approx$ 1 x 10$^4$ -- 5 x 10$^5$ km, 100 -- 800 km s$^{-1}$, and 10$^3$ -- 10$^5$ km, respectively \citep{Chandra2015,Sterling2016,Joshi2017}.  \citet{Raouafi2016} provides a comprehensive review of the coronal jet phenomena, including observations, theory, and numerical simulations.

According to the eruption process, \citet{Moore2010} classified solar jets in two sub-classes, {\it i.e.} standard and blowout jets. In a standard jet, the core field of the base arch remains close and static whereas in a blowout jet it explodes and results in a breakout eruption.
They further clarified that about two-third of the observed X--ray jets fall in the standard picture of jets and one third are of blowout category. The dichotomy of coronal jets into two categories is a result of the shear/twist in the base arch of the jet. Blowout jets usually have a high shear/twist in the base to erupt and open \citep{Liu2009, Chandra2017}. Helicity can be transferred from the closed field into the open field due to the reconnection between them. This ejection of helicity causes the motion of the jet material upwards by nonlinear torsional Alfv\`en waves \citep{Pariat2009, Jiajia2019}. The magnetic reconnection between the closed and open field lines is the causatum of magnetic flux
emergence and cancellation. The continuous magnetic flux cancellation and emergence destabilize the field at the jet base.

CMEs have attracted the solar physicists greatly as they are playing
a significant role in affecting the Earth's space environment. Usually CMEs are associated with large scale solar eruptions, {\it i.e.}, two--ribbon flares \citep{NJoshi2016, Zuccarello2017}, filament eruptions \citep{Schmieder2013,RChandra2017}, and occasionally with small scale solar eruptions, {\it i.e.}, solar jets \citep{Shen2012, Jiajia2015, Zheng2016, Sterling2018}. \citet{Shen2012} reported two simultaneous CMEs associated with a blowout jet. One of the two CMEs was bubble-like and the other was jet--like. The authors suggested that the external magnetic reconnection produced the jet--like CME and also led to the rise of a small filament underneath the jet base. Further, they explained that the bubble--like CME is due to the internal reconnection of the
magnetic field lines. \citet{Jiajia2015} observed a coronal jet event which led to a  high-speed CME (1000 km s$^{-1}$), suggesting that large--scale eruptions could be triggered by a small--scale jet. \citet{ Zheng2016} reported another similar event as a case study of solar jet activity which developed into a CME eruption.


However, the number of such jet--CME associated cases are too small for us to understand the mechanism and kinematic processes behind the phenomenon. Here, we present a jet event followed by a CME on 2013 April 28, which provides evidence of clear association of the jet and the CME. The jet erupted with an initial speed of $\approx$ 200 km s$^{-1}$ and developed into a CME together with the ambient coronal structures. The paper is structured as follows: we present the data analysis in section \ref{data}. The kinematics of jet and CME is given in section \ref{kinematics}. Magnetic field configuration of the jet
source region is described in section \ref{magnetic}. Finally, we discuss and summarize our results in section \ref{result}

\section{Data}\label{data}

The observational data for the jet eruption and the CME is taken from \emph{Solar Dynamics Observatory} \citep[SDO,][]{Pesnell12}, \emph{Solar 
TErrestrial RElations Observatory} \citep[STEREO,] [] {Kaiser2008}), and Large-Angle Spectrometric Coronagraph \citep[LASCO,] [] {Brueckner1995} onboard 
\emph{SOlar  and Heliospheric Observatory} \citep [SOHO,] [] {Domingo1995}. 
Atmospheric Imaging 
Assembly \citep[AIA,] [] {Lemen2012} onboard \emph{SDO} observes the 
Sun in seven EUV/UV wavelengths with a spatial resolution of 0.6 arcsec and a cadence 
of 12 seconds. For the multi-thermal jet structure, we analysed the AIA data in 131 \AA, 171 \AA, 
193 \AA, 211 \AA, and in 304 \AA. For a better contrast of the hot and cool counterparts of 
the jet, we create the base and running difference images of the AIA data.
To probe the jet and CME from multiple perspectives the EUV images taken by
the Sun Earth Connection Coronal and Heliospheric Investigation (SECCHI \citep{Howard2008}) 
onboard \emph{STEREO} are analyzed. The twin spacecraft of \emph{STEREO} mission, \emph{STEREO}--A and 
\emph{STEREO}--B observe the Sun from two angles in four different EUVI channels, 171 (Fe IX), 
195 (Fe XII), 284 (Fe XV), and in 304 (He II) \AA. For our current analysis of the jet, 
we use the EUV images of \emph{STEREO}--B in 304 \AA\ with a cadence of 10 minutes and pixel size of
1 arcsec. \emph{STEREO}--A and B were separated by 83$^{\circ}$ on 2013 April 28.
To correct the projection effect for the speed of jet, we use SCC$_{-}$MEASURE routine of 
SECHHI available in SolarSoft library in IDL. In this routine``tiepointing'' technique is used to 
reconstruct the three dimensional picture of the ejecting feature, by clicking the same 
feature on both \emph{STEREO} and \emph{SDO} images \citep{Thompson2006, Gosain2009}.

The CME is well observed with  {\it SOHO}/LASCO and \emph{STEREO}/COR coronagraphs. \emph{STEREO} COR1 has a field of view from 1.5 to 4 R$_{\odot}$ and provides the images with a cadence of 5 minutes, while COR2 observes the corona from 2 to 15 R$_{\odot}$. The LASCO observed 
the CME in the outer corona upto 30 R$_{\odot}$ with a cadence of 12 (C2) and 30 (C3) minutes. With the multi-point observations from LASCO and COR, we employ the Graduated Cylindrical Shell (GCS) model to obtain the three--dimensional height and direction of the CME (see section \ref{cme} ). We further analyse the photospheric magnetic field using  the line-of-sight magnetograms from  Helioseismic and Magnetic Imager \citep[HMI,] [] {Schou2012} onboard \emph{SDO}. 
For a closer and clear view of the jet source region, we use HMI Spaceweather HMI Active Region Patch \citep[SHARP,] [] {Bobra2014} data set with a cadence of 12 minutes.
\begin{figure*} 
\centering
\includegraphics[width=0.9\textwidth]{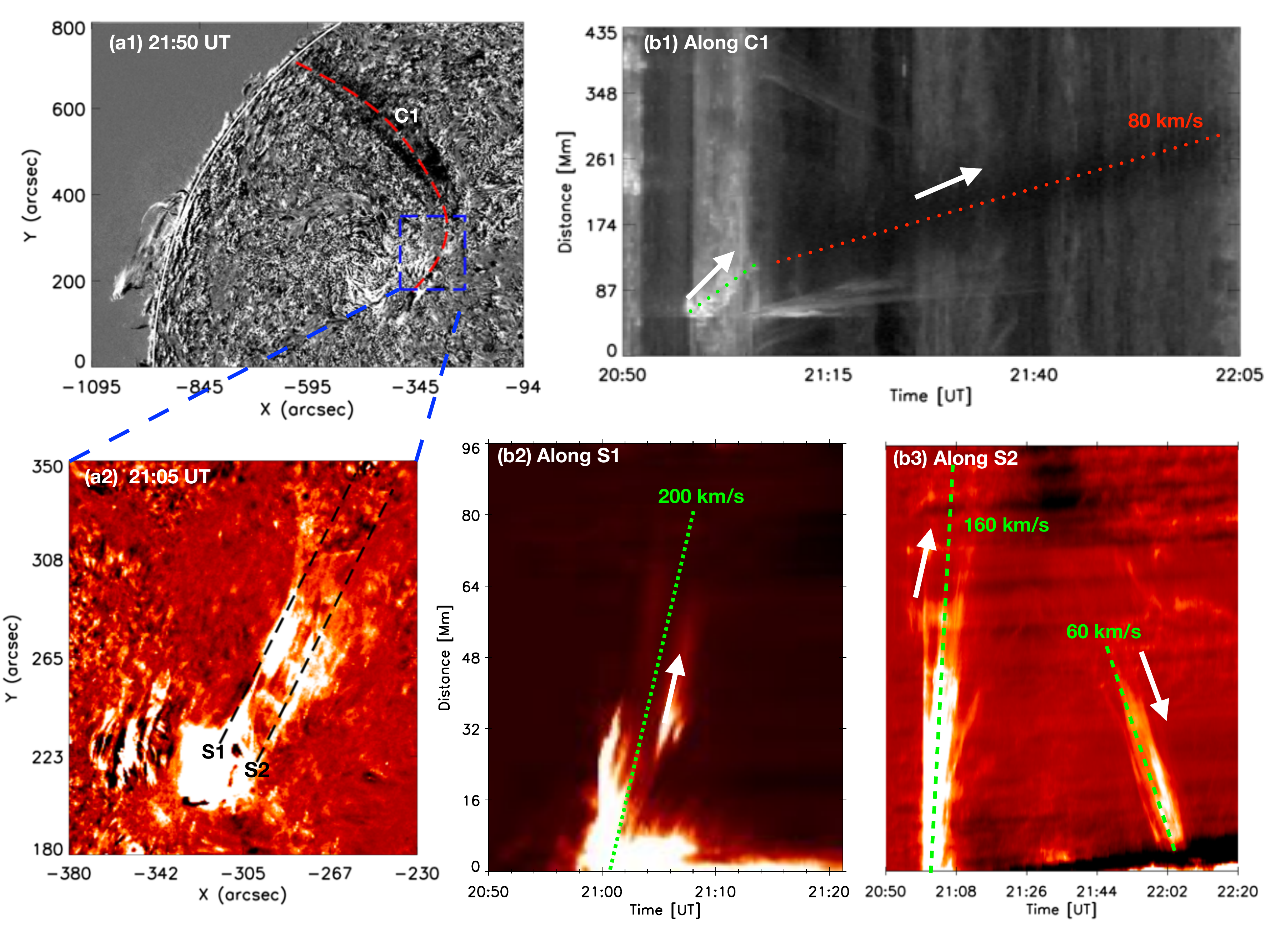}
\caption{The observed jet (a1-a2) and height--time analysis (b1-b3) with AIA 304 \AA\ on 2013 April 28 from NOAA AR 11731. Panel (a1) shows the direction of jet propagation from the 
source region (blue rectangular box) towards the solar limb along with the red dashed curve and 
(b1) is the space time plot along the curve C1. The jet starts with a speed of $\approx$ 200 km s$^{-1}$.
Panel (a2) is the location for the jet source region which is indicated by the blue rectangular box 
in panel (a1). Panel (b2) and (b3) are the height--time analysis along slit S1 and S2 respectively, 
indicated in panel (a2). In the direction of slit S2, the jet material falls back into the source region 
with a speed of $\approx$ 60 km s$^{-1}$.
An animation (MOV1) of panels a1 and b2) is available, which presents the jet eruption from its base towards the eastern limb from 20:50 UT to 21:57 UT. 
}
\label{timeslice}
\end{figure*}
\newpage
\section{Kinematics}\label{kinematics}
\subsection{Kinematics of the Solar Jet}

The jet started to erupt $\approx$ 20:53 UT with a circular base, towards the northern direction from the AR NOAA 11731 (N09E23) and observed in all six AIA channels (94 \AA, 131 \AA, 171 \AA, 193 \AA, 211 \AA, and 304 \AA). After reaching to some height at about 80 Mm, the jet material was deflected from its original direction of propagation and revolved around the north--east direction. The jet was initially bright (Figure \ref{timeslice}(a2)), and afterwards followed with dark material (Figure \ref{timeslice}(a1)), suggesting impulsively strong heating at the initial phase. The following dark material was only visible in AIA 304 \AA\ and not observed in hot channels, {\it i.e.} 171 \AA. The propagation of the whole jet in AIA 304 \AA\ is shown in Figure \ref{timeslice}(a1) along with the red curve C1, which indicates the deflection of the jet from north direction to north--east direction towards the solar limb. The initiation of the jet from the source region is shown in panel (a2). We also observed a small jet ejection at about 21:24 UT in the eastern neighbourhood of the source region, and this jet material merged with the big jet (animation is attached with Figure \ref{timeslice} as MOV1). Panel (b1) is the height--time plot of the jet along the slit C1. The jet speed shows a two--stage profile. The speed in the later stage is about 80 km s$^{-1}$ towards the north--east direction (red dotted line). For the velocity in the initial stage, we set two slits S1 and S2 (panel (a2) of 10 pixel width in two different directions, and found that the speed in the S1 direction is 200 km s$^{-1}$ and that the other direction S2 is about 160 km s$^{-1}$
(as presented in panel (b2) and (b3)). In addition to this, we found
that a portion of the jet material falls back to the source region around 21:51 UT with a speed of $\approx$ 60 km s$^{-1}$, clearly appeared in height--time plot along S2 direction in panel (b3).

\begin{figure*} 
\centering
\includegraphics[width=0.8\textwidth]{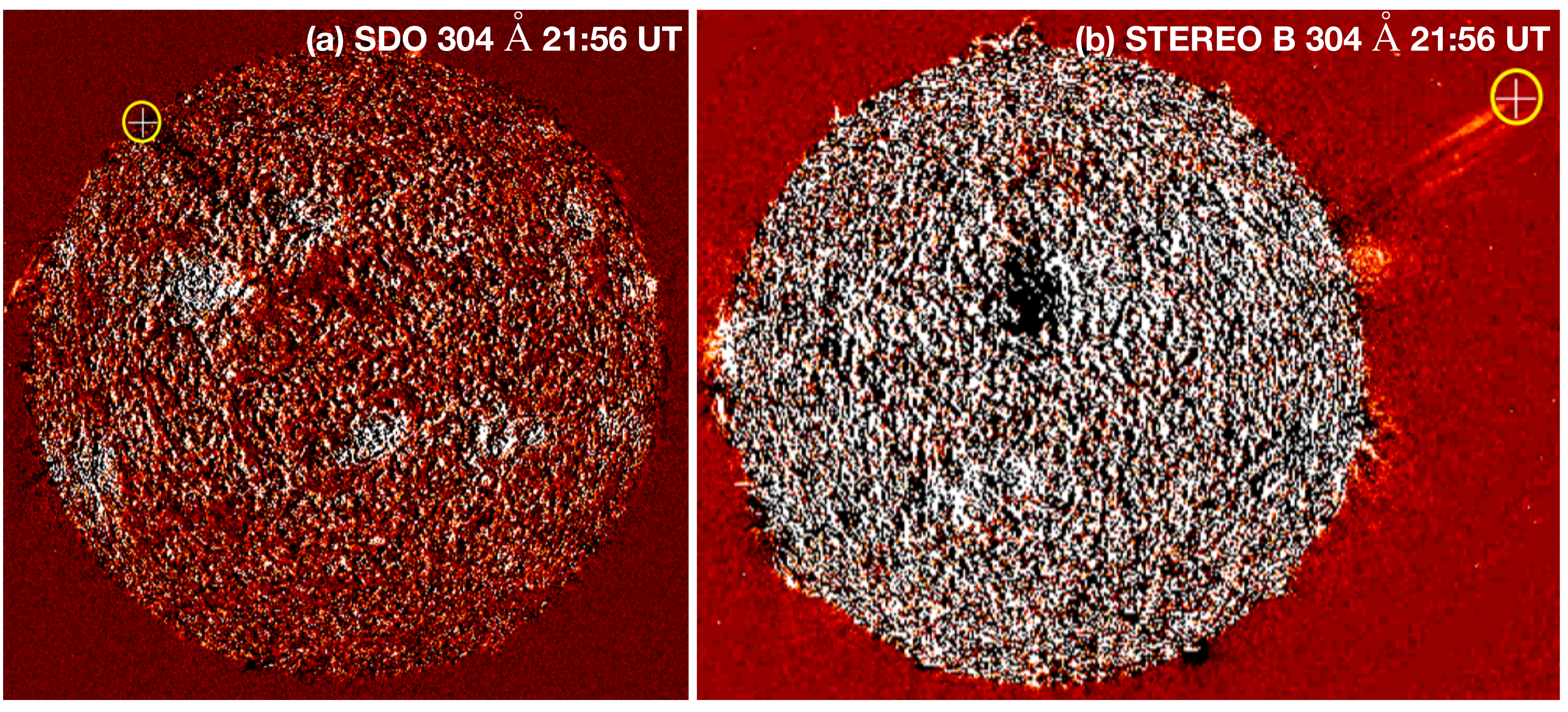}
\caption{The leading edge of the jet material in \emph{SDO}/AIA 304 \AA\ (a) and in \emph{STEREO}--B EUV 304 \AA\ (b).
 The $+$ sign encircled with yellow circle shows the 
location of the leading edge of the jet at 21:56 UT. These two positions of the leading edge of jet are obtained from the SCC$_{-}$MEASURE technique which is used for the jet velocity calculation.}
\label{scc}
\end{figure*}

From  $\approx $ 21:16 UT, \emph{STEREO}--B observed the cool counterpart of a jet in 304 \AA\ above the western limb.
The full-disk image of AIA 304 \AA\ and \emph{STEREO}--B EUV 304 \AA\ is presented in Figure \ref{scc}. The highest visible peak of the jet is indicated with a circle at the solar limb which is used to get the read jet speed. Figure \ref{allcme} (panel (c)) showed the locations of \emph{STEREO} satellite, the Earth and the Sun. With the aid of SCC$_{-}$Measure procedure, we get the real speed and propagation direction of the jet by clicking on the same feature in AIA 304 \AA\ and in \emph{STEREO}--B 304 \AA\ image. The real jet-speed was 200 km s$^{-1}$ towards the north--east (longitude = -18$^{\circ}$, latitude = 19$^{\circ}$) direction. However, this correction can be only applied to the second stage of the jet when it was propagating towards the north--east direction, because we do not have the stereoscopic observations for the early stage of the jet.

\subsection{Kinematics of the  CME}
\label{cme}
\begin{figure*} 
\centering
\includegraphics[width=0.60\textwidth]{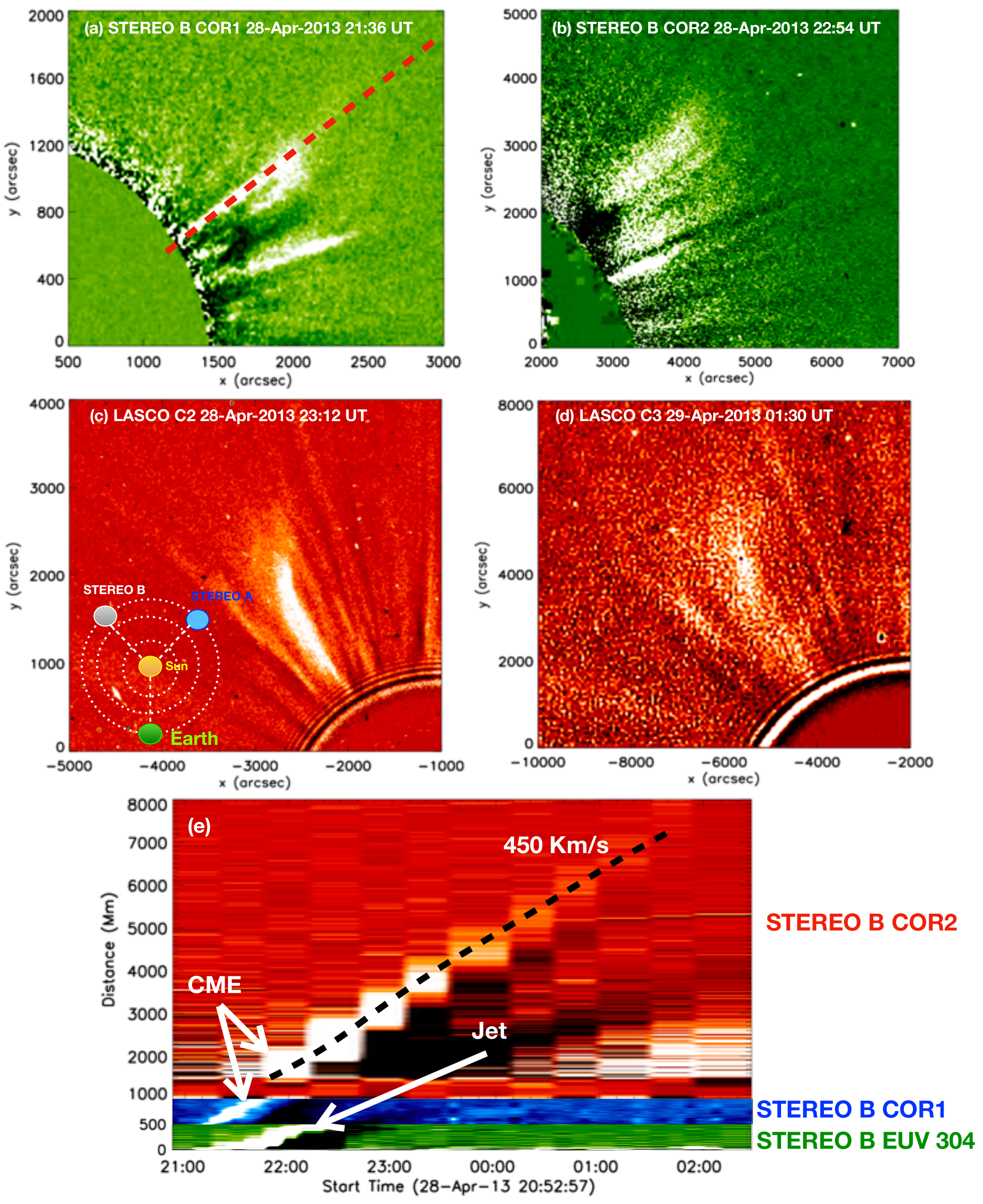}
\caption{CME associated with the jet is observed by LASCO and \emph{STEREO} corongraphs.
Top: Observed CME in \emph{STEREO}--B COR1 and COR2. Middle: Observed CME in LASCO C2 and C3 field of view. The location of \emph{STEREO} A, B, and the Earth
on 2013 April 28 is 
shown in panel (c).
Bottom: Height--time analysis of the jet and jet--like CME observed with \emph{STEREO}--B EUVI 304 \AA (green), COR1 (blue), and COR2 (red). The direction of the slit for this height--time analysis is shown in panel (a) with red dashed line. The separation
between the CME front and jet front is due to the expansion of the jet--like CME. They almost originated from the same location when extrapolated back to the solar surface. An animation for the jet-CME association with \emph{STEREO}--B instrument is available as MOV2. The animation shows the jet eruption observed with \emph{STEREO}--B EUV 304 \AA from 2013 April 28 20:36 UT to 2013 April 29 00:06 UT, \emph{STEREO}--B COR1 from 2013 April 28 21:00 to 22:10 UT, and \emph{STEREO}--B COR2 from 2013 April 28 20:54 UT to 2013 April 29 03:24 UT.
}
\label{allcme}
\end{figure*}

The associated CME was well observed with {\it SOHO}/LASCO and \emph{STEREO}--B COR1 and COR2 coronographs,  as shown in Figure \ref{allcme}. The CME is not a typical one. It is narrow (width $\approx$ 25$^{\circ}$), and likes a giant jet in the corona, no matter from which perspective the CME was viewed. The jet--CME association is very much evident in \emph{STEREO}--B observations (see Figure \ref{allcme}(e) and the attached animation MOV2). For the continuous tracking of the solar jet in EUV channel (304 \AA) and the CME in coronagraphs, we put a slit in the jet--CME direction in \emph{STEREO}--B EUV 304 \AA, COR1 and COR2. The direction of the slit is shown in Figure \ref{allcme}(a). The continuous spatial and temporal correlation between the jet and the CME is presented in Figure \ref{allcme}(e).  The front of the CME is much higher than the jet front and the separation
between them is due to the expansion of the CME, causing the speeds of their fronts are different. If extrapolating them back to the solar surface, they almost originated from the same time.

\begin{figure*} 
\centering
\includegraphics[width=0.6\textwidth]{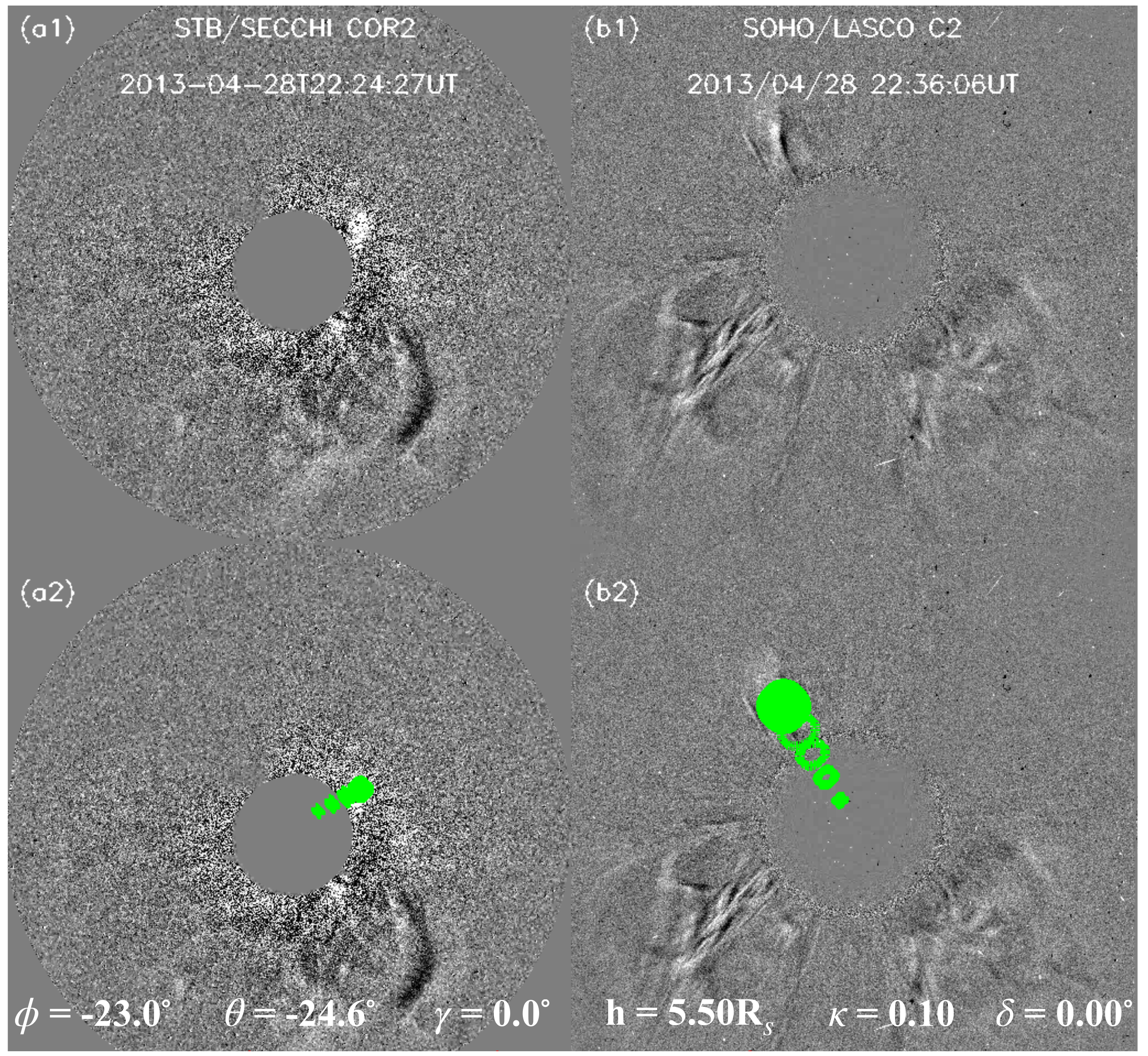}
\caption{CME associated with the jet eruption is analysed with 
 GCS model, which indicates the direction of propagation of the CME 
in \emph{STEREO}--B COR2, and LASCO C2 with a speed of $\approx$ 450 km s$^{-1}$.}
\label{gcs}
\end{figure*}
To reduce the projection effect, we use the GCS model to get the real kinematic properties of the CME. The GCS model is developed to represent the flux rope structure of CMEs \citep{Thernisien2006, Thernisien2011}. It involves three geometric parameters: `h', the height of the leading edge, `$\kappa$', the aspect ratio, and `$\delta$', the half edge-on angular width, and three positioning parameters: `$\theta$', `$\phi$', and `$\gamma$', the Stonyhurst latitude and longitude of the source region, and the tilt angle of the source region neutral line respectively. The GCS model is usually used to study morphology, position, and kinematics of a CME based on the best fitting result of a CME transient recorded in white-light images. The ice-cream cone model is another model of CMEs, which composed of a ball that we call the ice-cream ball and circular cone tangent to the ball with a conic node on the solar surface \citep{Fisher1984}. The GCS model becomes equivalent to the ice-cream cone model when its parameter $\delta$, equals 0 \citep{Thernisien2011}. For our case study, we use the ice-cream cone model which is a simplified form of the GCS model and estimated the three-dimensional height and direction of the CME with LASCO C2, C3 and \emph{STEREO}--B COR2 images. The best-fitted GCS model is  displayed in Figure \ref{gcs}. The corrected CME speed from the GCS model comes out to be 450 km s$^{-1}$.

\begin{figure*} 
\centering
\includegraphics[width=1.0\textwidth,angle=0]{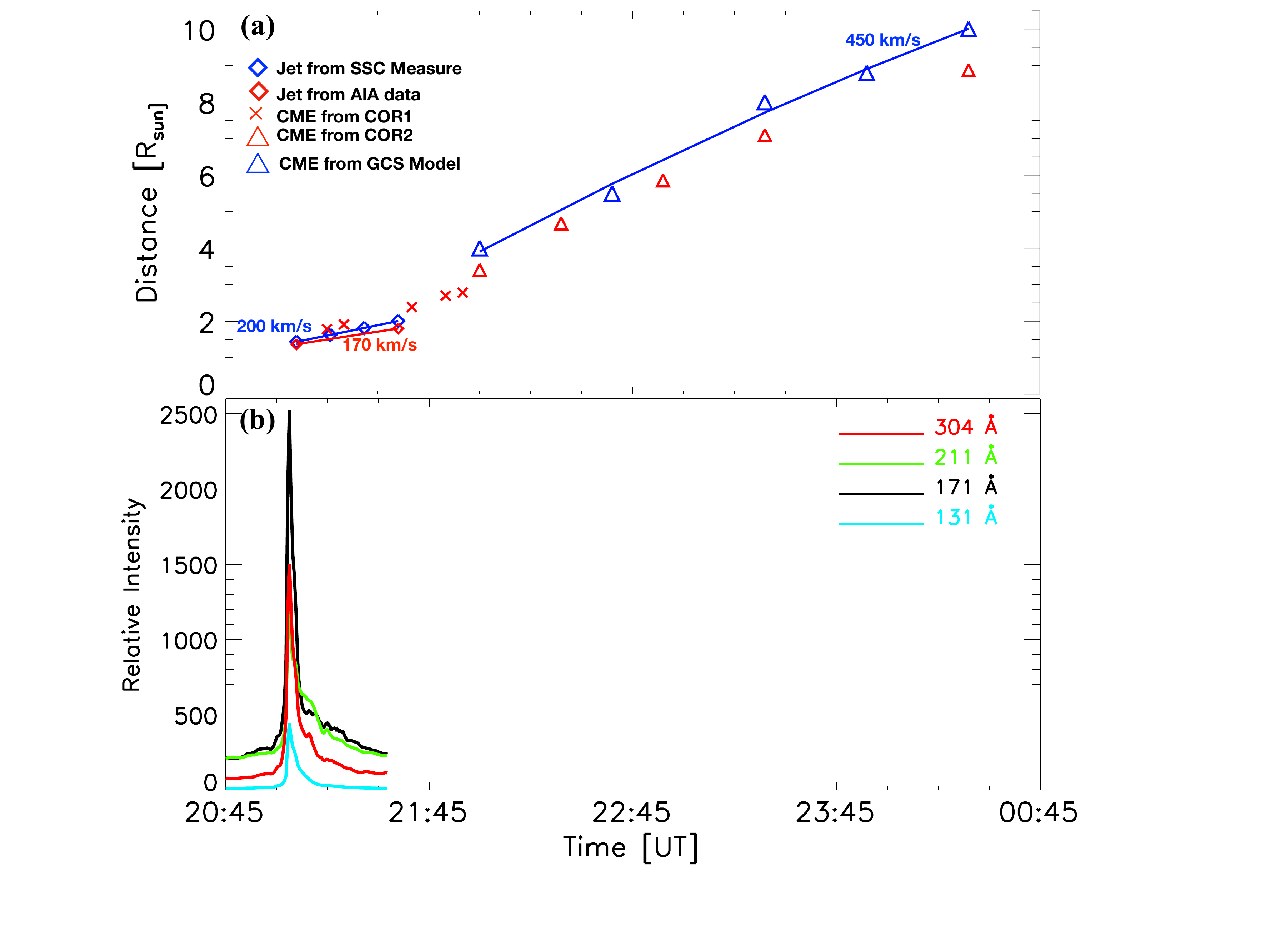}
\vspace*{-1.5cm}
\caption{
Panel (a): The complete kinematics of the jet and the CME. The projection corrected speeds are plotted 
with blue color, while the red points are used for uncorrected data. Panel (b): Light curves for different wavelengths.
These are observed at the jet base shown as the red rectangular box presented in Figure \ref{hmi}(e).}
\label{lightcurve}
\end{figure*}
Figure \ref{lightcurve}(a) depicts the complete kinematics of the jet and the CME with the different data points of various instruments.
We have corrected the projection effect for the jet and CME with SCC$_{-}$MEASURE and GCS model fitting, respectively. The corrected jet speed comes out to be $\approx$ 200 km s$^{-1}$ from SCC$_{-}$MEASURE associated with a CME of speed $\approx$ 450 km s$^{-1}$. The blue and red colors are used for corrected and uncorrected data points. This plot of temporal evolution shows the clear link between the jet and the narrow CME. Figure \ref{lightcurve}(b) shows the intensity
variation at the jet base. The impulsive peaks at the jet base show the jet peak time in various AIA wavebands. The enhancement in the light curve of EUV emission suggests that the energy injection was at the very beginning only, and not responsible for the continuous acceleration of the jet to escape from the Sun.

The speed of the CME (450 km s$^{-1}$) is much larger than that of a jet (200 km s$^{-1}$). This is because the speed of different parts of erupting structures are measured.
The speed of CME obtained from the \emph{STEREO} and LASCO observations 
is at its leading edge (v$_{front}$). It consists of the propagation speed of the CME  center (v$_{center}$) and the expansion speed (v$_{exp}$) of the CME, so v$_{front}$ = v$_{center}$ + v$_{exp}$. A cartoon illustrating the CME speed at the leading edge, which includes the CME propagation speed and expansion speed is given in \cite{Yuming2015} (see their Figure 1).
\citet{Gopalswamy2009} derived a relation between CME propagation speed
and expansion which is confirmed in many studies till now
\citep{Michalek2009,Makela2016}. With an approximation of the CME shape
by a shallow ice cream cone, the relationship is defined as v$_{exp}$ =
2 v$_{front}$ sin$(w/2)$, where `w’ is the CME width (25$^{\circ}$ in
present case). Therefore v$_{exp}$ comes out to be 230 km s$^{-1}$ and
v$_{center}$ should be 220 km s$^{-1}$. The jet triggered and
developed into the CME and its trajectory should be followed by the CME
center and not by the leading edge of the CME. Thus, the jet velocity
(200 km s$^{-1}$) is comparable with v$_{center}$ (v$_{center}$ $<$
v$_{front}$). That explains the difference between the jet and CME
speeds.


\begin{figure*} 
\centering
\includegraphics[width=1.00\textwidth]{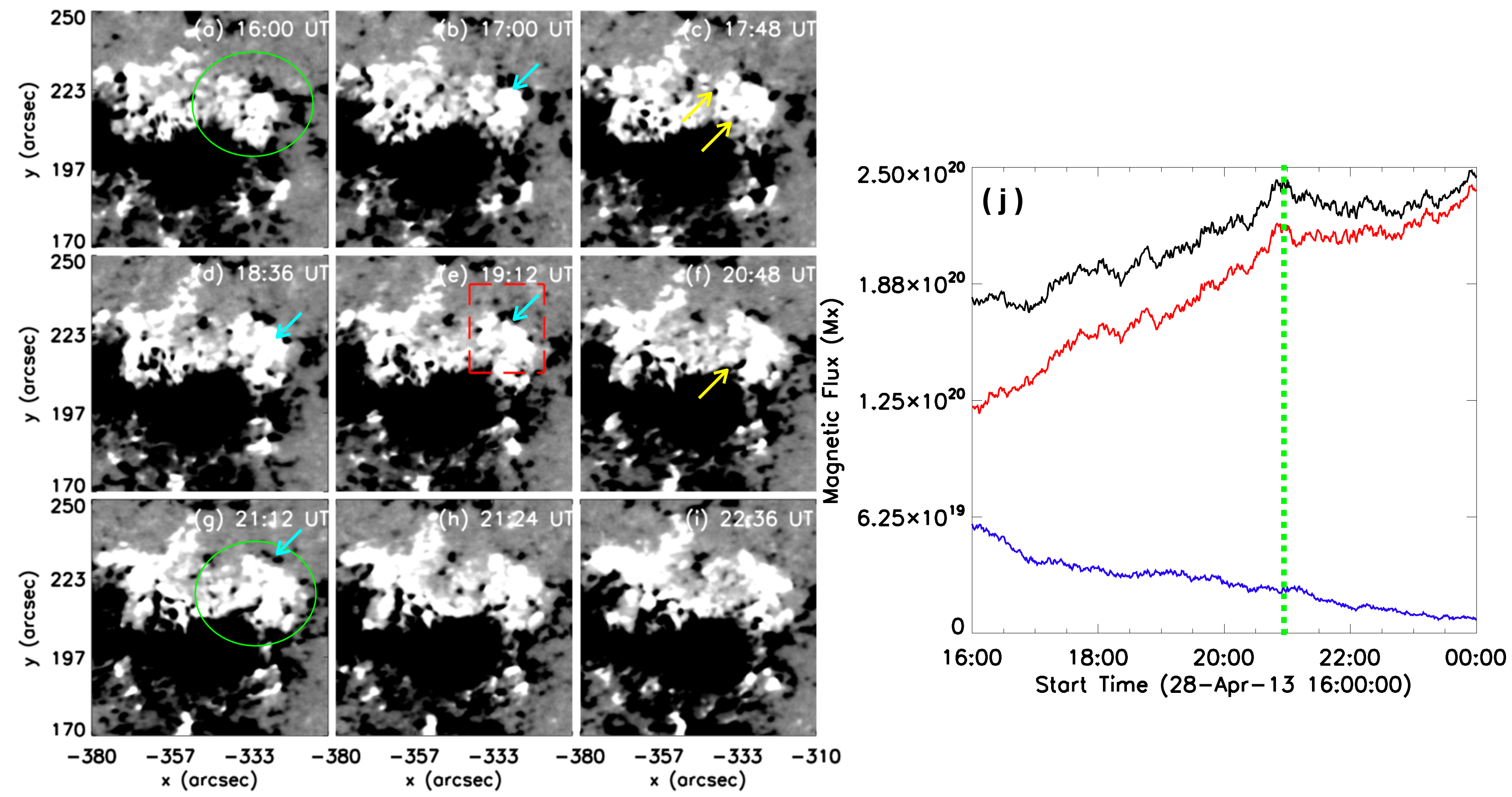}
\caption{
The left panels (a)--(i) show the magnetic field configuration at the jet site. Cyan and 
yellow arrows show the cancellation and emergence of negative magnetic polarity. Emergence of 
positive magnetic polarity is enclosed by green circle in panel (a) and (g). Right panel (j) is 
the magnetic flux variation with time calculated at the jet source region indicated as red rectangular box in the left panel (e). Jet starting time is highlighted with a green dashed line.
Black, red, and blue curves are for the total, positive, and negative magnetic flux respectively.
An animation for the magnetic field evolution (a-i) is available from 16:00 UT to 22:36 UT.}

\label{hmi}
\end{figure*}
\section{Magnetic Configuration of the Source Region}\label{magnetic}

For a better understanding of the trigger mechanism of the solar jet, we did the magnetic field analysis of the source region using the HMI SHARP data (12 min cadence) of the AR 11731 on 2013 April 28. The continuous cancellation of the negative magnetic polarity by the emerging
positive magnetic spot is observed (Figure \ref{hmi} and accompanying animation MOV3). The positive magnetic polarity ate the negative magnetic polarity which was already distributed in the jet source region (plotted inside the green circle in the panel (a) and (g)). Afterwards, small negative polarities emerges from the large negative ball and get cancelled with the big positive polarity area. The emergence of small negative polarities is shown with yellow arrows and the cancellation is indicated with cyan arrows. To look at the variation of the magnetic flux with time, we calculated the positive, negative and total unsigned magnetic flux at the jet source region, which is indicated as the red rectangular box in panel (e). This is the same dimensional area we used to calculate the light curve in Figure \ref{lightcurve}(b). The flux variation with time in panel (j) shows that, there is a continuous cancellation and emergence of the negative magnetic flux (blue line) while the positive magnetic flux emerges throughout (red curve). The total unsigned magnetic flux (black curve) shows the simultaneous cancellation and emergence of magnetic polarities at the jet source region. The emergence of the positive magnetic flux dominated over the cancellation throughout. The initiation of jet time is shown with a green dashed vertical line.

We analysed the magnetic topology at the jet location and applied two different methods of  potential extrapolation, one for the global overview of the jet eruption, and the other for the local view at the jet base region. The details are as follows:

We apply the Potential Field Source Surface (PFSS) model \citep{Schrijver03}, to investigate the global magnetic topology  near the jet source region. This PFSS technique uses the HMI synoptic magnetic maps processed with a  software package available in SSWIDL. 
We apply the PFSS technique to see the reconnection between close loops and open field lines because at large scale the corona is in potential state \citep{BSchmieder1996}. The PFSS model for this case study is presented  in Figure \ref{pfss}(a), with open (pink) and close (white) magnetic field lines. These open field lines 
resemble exactly the path exactly the same as the jet, which was along the north direction in the beginning and deflected  towards the north--east afterwards.

To describe the magnetic topology of the jet base region, we extrapolate the coronal potential field  using the photospheric  Line of Sight (LOS) magnetogram as a boundary condition. The method based on the   Fourier transformation (FT) method proposed by \citep{Alissandrakis1981}. The FT method requires the vertical component of the photospheric vector field as the input parameter. However, due to  HMI vector magnetic field  limited field of view, the extrapolation hard to meet divergence--free condition. Hence, we cut a larger patch of the LOS magnetogram instead. As the AR is close to the central meridian, the LOS magnetic field could  represent the vertical field to a large extent. In the extrapolated magnetic field, we find open field lines coincide well with the extension direction of the jet shown in Figures \ref{pfss}b and c.

  
\section{Discussion and Summary}\label{result}

In this article, we investigate a solar jet eruption from the AR
NOAA 11731 on April 28, 2013. The jet was ejected towards the north direction, and after reaching some height (80 Mm), it was deflected in the north-east direction. The average computed speed of the jet was $\approx$ 200 km s$^{-1}$. We found a clear association of the observed
jet with a narrow CME of speed 450 km s$^{-1}$, observed in LASCO C2, C3 and \emph{STEREO}--B COR1, and COR2 coronagraph.

\begin{figure*} 
\centering
\includegraphics[width=1.0\textwidth]{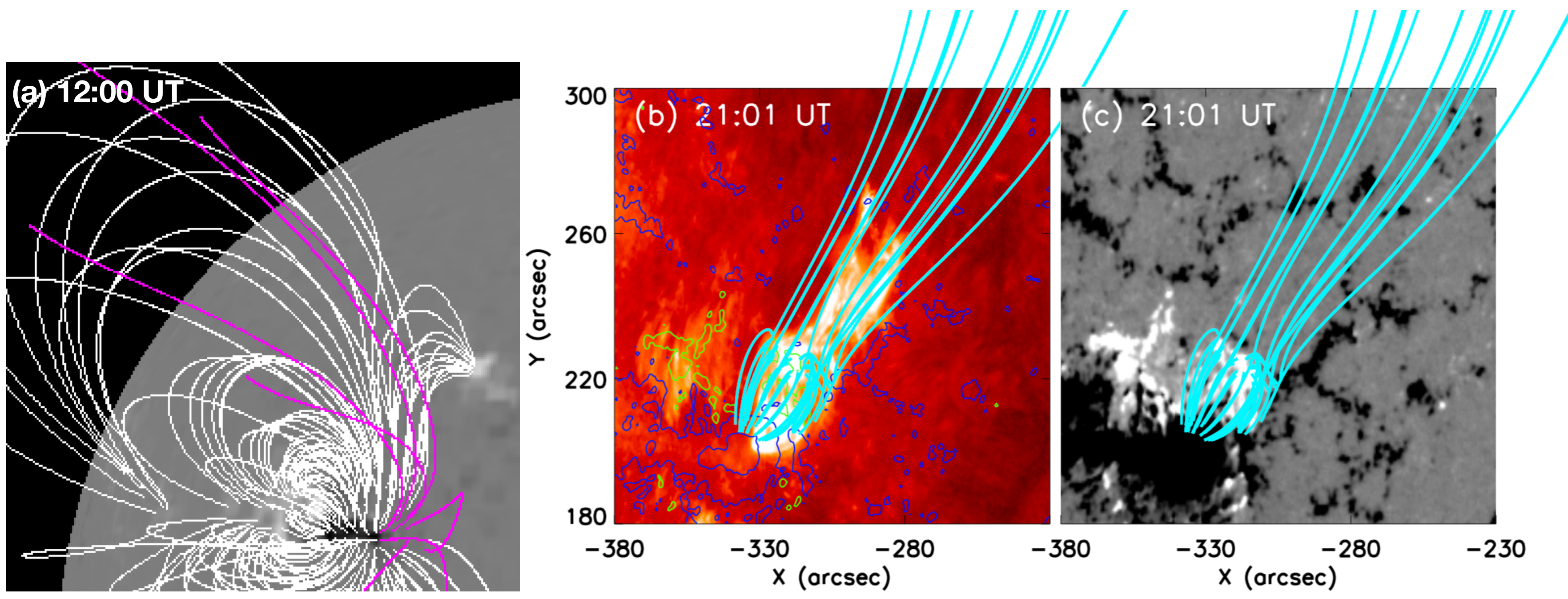}
\caption{The PFSS extrapolation of the large FOV is shown in panel (a). The white and pink lines are 
the closed and open magnetic field lines at the jet location. The open field lines are resembling in 
the same way as of the jet propagation from its source to solar corona, which is indicated as curve 
C1 in figure \ref{timeslice}(a1). Panel (b) (AIA 304 \AA) and (c) (HMI magnetogram) are showing the
jet source region with the same FOV as of the blue rectangular box in figure \ref{timeslice}(a1). 
Cyan lines from the source region  are the magnetic field lines, which shows closed structure at the 
jet base and open lines afterwards. Blue and green contours in panel (b) are of negative and positive  magnetic field.}  
\label{pfss}
\end{figure*}
The escape speed of Sun is given as $v = \sqrt{\frac{2GM_\odot}{R}}$, where R is the distance from the center of the Sun. The observed jet speed computed using the multi-view point observations is about 200 km s$^{-1}$ at the height of 2 R$_\odot$. The escape velocity
computed at the height of 2 R$_\odot$ using the above formula is $\approx$ 430 km s$^{-1}$. Therefore, we conclude that the complete jet cannot be escaped from the solar surface. This could be the reason we have observed the backward motion of the jet material from the propagation direction towards the source region. Even the jet speed is lower than the escape speed, we observed the clear CME associated with the jet by all the space-borne coronagraphs. The possible mechanism for the jet continuously accelerating to reach the escape speed and form the narrow CME is that the falling back material makes the upward material of the jet moving faster to keep the momentum of the whole jet conserved. We concluded that the observed speed of the CME is containing the speed of the CME center and the expansion speed, and is much larger than the jet speed, because the different parts of the erupting structures are being measured. The speed of CME center (the trajectory followed by the jet) is 220 km s$^{-1}$ and  is equivalent to the speed of the jet (200 km s$^{-1}$). This provides a clear evidence of the jet-CME association.

For the magnetic configuration at the jet origin site, two views are popular. One is the magnetic flux emergence observed in many observations and also proposed in the MHD simulations \citep{Shibata1992, Yokoyama95, Moreno2008, Moreno2013, Ruan2019,Joshi2020}. Another is the magnetic flux cancellation, which is also reported in the observations MHD simulations \citep{Pariat2009, Young2014, Chandra2017, McGlasson2019}.
We have also studied the magnetic field evolution in our investigation and observed that there is a continuous emergence and cancellation of the negative magnetic flux and the positive flux is emerging throughout. Therefore, we believe that both the flux emergence and the cancellation are responsible in our case.

In the potential field extrapolation, we have found that the jet source region is covered (or overlaid) by the closed field lines and open field lines. The open field lines are in the direction of the jet ejection.
We have tried for the non-linear force-free field (NLFFF) extrapolation, but it failed to reproduce the magnetic field topology of the AR. The field lines of the NLFFF did not resemble with the loops observed in the EUV passbands. It might because that the FOV of the photospheric vector magnetic field provided by HMI SHARP data is too small as it is available for significant AR patch of solar magnetic field. Hence the divergence-free condition is not completely satisfied in the extrapolation, which makes the NLFFF results unreliable. On the other hand, we mainly focus on the propagation of the jet, which is more likely to be relevant to the nearly potential, large scale magnetic field connectivity. Therefore we believe that the potential field extrapolation might be sufficient and the direction of the jet ejection is the same as of the
open magnetic field lines we have obtained from the potential field extrapolation (see Figure \ref{pfss}). Hence we conclude that the reconnection between the closed and open magnetic field lines provides
a path for the ejection of the jet.


We also observed the rotation in the jet material when it is propagating towards the north direction 
from the source region (animation is attached as MOV1). The untwisting of the jet suggests the injection of helicity to the upper atmosphere.

In summary, our study of a narrow CME caused by the jet gives worth evidence that some jets from the solar disk can escape from the corona to form a CME, which may contribute to the space weather phenomena. For our future study, we are looking forward to  finding the clear {\it in situ} measurements for such jet--like CMEs from the newly launched Parker Solar Probe.

{\bf Acknowledgments}\\
We thank the anonymous referee for his/her valuable comments which improve our manuscript to a significant level. We thank \emph{SDO}, \emph{STEREO}, and \emph{SOHO} science teams for the free access to the data. RJ thanks to 
SCOSTEP Visiting Scholarship program for providing the opportunity to carry this work at University of Science and Technology China (USTC) and 
to the Department of Science and Technology (DST), New Delhi India, for the INSPIRE fellowship. RJ also thanks to Dr. Arun K. Awasthi for the useful discussion during her visit at USTC. This work is supported by the grants from NSFC (Nos. 41774178 and 41761134088). RC acknowledges the support from Bulgarian Science Fund under Indo-Bulgarian bilateral project. We thank Dr. Brigitte Schmieder for helpful discussion.


\end{document}